\documentclass[%
twocolumn,
superscriptaddress,
nofootinbib,
longbibliography,
amsmath,amssymb,
aps,
]{revtex4-2}
\usepackage{pdfpages}
\makeatletter
\AtBeginDocument{\let\LS@rot\@undefined}
\makeatother
\usepackage{graphicx}
\usepackage{bm}
\usepackage{dsfont}
\usepackage{multirow}
\usepackage{mathrsfs}
\usepackage{leftidx}
\usepackage{xspace}
\usepackage{diffcoeff}  
\usepackage{bbm}
\usepackage{amsfonts,amssymb,amsmath}
\usepackage{mathtools}

\usepackage{array} 
\usepackage{tabularx}
\usepackage{bbold}
\usepackage{pifont}
\usepackage[normalem]{ulem}

\usepackage[a4paper]{hyperref}
\hypersetup{colorlinks=true,linktoc=all,linkcolor=blue,breaklinks=true,citecolor=blue,urlcolor=blue}
\usepackage{siunitx}

\usepackage[utf8]{inputenc}
\usepackage[english]{babel}

\addto\captionsenglish{}

\AtBeginDocument{%
    \newwrite\bibnotes
    \def\bibnotesext{Notes.bib}
    \immediate\openout\bibnotes=\jobname\bibnotesext
    \immediate\write\bibnotes{@CONTROL{REVTEX42Control}}
    \immediate\write\bibnotes{@CONTROL{%
    apsrev42Control,author="08",editor="1",pages="1",title="0",year="1"}}
     \if@filesw
     \immediate\write\@auxout{\string\citation{apsrev42Control}}%
    \fi
}%

\newcommand{\be}{\begin{equation}}
\newcommand{\ee}{\end{equation}}
\newcommand{\beq}{\begin{eqnarray}}
\newcommand{\eeq}{\end{eqnarray}}

\begin{document}
\title{
Nonlocal thermoelectric detection of interaction and correlations in edge states}
\author{Alessandro Braggio}
\affiliation{NEST, Istituto Nanoscienze-CNR and Scuola Normale Superiore, Piazza San Silvestro 12, I-56127 Pisa, Italy}
\author{Matteo Carrega}
\affiliation{SPIN-CNR, Via Dodecaneso 33, 16146 Genova, Italy}
\author{Bj\"orn Sothmann}
\affiliation{Theoretische Physik, Universit\"at Duisburg-Essen and CENIDE, D-47048 Duisburg, Germany}
\author{Rafael S\'anchez}
\affiliation{Departamento de F\'isica Te\'orica de la Materia Condensada, Condensed Matter Physics Center (IFIMAC), and Instituto Nicol\'as Cabrera, Universidad Aut\'onoma de Madrid, 28049 Madrid, Spain\looseness=-1}
\date{\today}

\begin{abstract}
We investigate nonequilibrium effects in the transport of interacting electrons in quantum conductors, proposing the nonlocal thermoelectric response as a direct indicator of the presence of interactions, nonthermal states and the effect of correlations.  
This is done by assuming a quantum Hall setup where two channels (connected to reservoirs at different temperatures) co-propagate for a finite distance, such that a thermoelectrical response is \emph{only} expected if the electron-electron interaction mediates heat exchange between the channels. This way, the nonlocal Seebeck response measures the interaction strength. Considering zero-range interactions, we solve the charge and energy currents and noises of a nonequilibrium integrable interacting system, determining the universal interaction-dependent length scale of energy equilibration. Further, a setup with two controllable quantum point contacts allows thermoelectricity to monitor the interacting system thermalisation as well as the fundamental role of cross-correlations in the heat exchange at intermediate length scales.
\end{abstract}

\maketitle

\emph{Introduction.}--
The thermoelectric response of nanoscale devices~\cite{benenti:2017} provides a unique tool to probe quantum phenomena such as electron entanglement \cite{sanchez_cooling_2018,Hussein2019,Tan2021}, superconducting~\cite{hwang_odd_2018,claughton_TE_1996,germanese_BTE_2022} and topological properties~\cite{Takahashi2012Nov,ronetti:2016,erlingsson_reversal_2017,Xu2017Sep,blasi_nonlocal_2020}, entropy~\cite{kleeorin:2019} or photon-assisted tunnelling~\cite{hijano_pat_2023}. It can also serve as a probe of chirality in the quantum Hall regime~\cite{granger_observation_2009,nam_thermoelectric_2013,sanchez_chiral_2015}. The direct conversion of temperature differences into measurable electrical quantities allows for the natural exploration of heat exchange processes in quantum-coherent systems far from equilibrium~\cite{kovrizhin_equilibration_2011,tabatabaei_nonlocal_2022}.
Multiterminal configurations allow for a separation of heat-injecting and charge-propagating channels~\cite{thierschmann:2015,jaliel:2019,dorsch:2020}.

Quantum Hall edge channels constitute co-propagating electronic channels (ECs) which can be contacted electrically separately~\cite{fujisawa_review_2022}, thus allowing for precise electrical manipulation.
The role of electron-electron interactions has been investigated via the energy relaxation from nonthermal states injected either by a quantum point contact (QPC)~\cite{levkivskyi_energy_2012} or as hot electrons~\cite{ota_spectroscopic_2019,akiyama_ballistic_2019}. However, their detection is challenging~\cite{
hashisaka_waveform_2017,itoh_signatures_2018,rodriguez_relaxation_2020,suzuki_nonthermal_2023,konuma_nonuniform_2022}.
Moreover, thermal probes have been used to measure quantized thermal conductances with edge states~\cite{jezouin_quantum_2013}, to address the nature of edge states~\cite{kane_quantized_1997,Banerjee2018Jul,ma_equilibration_2019,breton_heat_2022,Dutta2022Jan}, to explore equilibration mechanisms~\cite{melcer_absent_2022,Srivastav_vanishing_2021} and to image thermal decay~\cite{arguerite_imaging_2019,moore_thermal_2023}.
\begin{figure}[b]
\includegraphics[width=\linewidth]{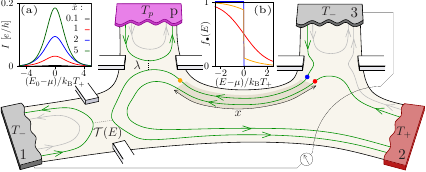}
\caption{\label{fig:qhsetup}
Quantum Hall detection setup. The interaction of the upper ECs along the distance $x$ before a scatterer of transmission probability $\mathcal{T}(E)$ enables a thermoelectric response between terminals 1 and 3, both at temperature $T_-$, when terminal 2 injects into the inner channel electrons at a temperature $T_+\neq T_-$. Inset (a) shows the thermoelectric current for interacting regions of different lengths ${x}$ and $T_-=0.5T_+$, when $\mathcal{T}(E)=\Theta(E{-}E_0)$ (e.g., a QPC), and (b) the distribution of the different channels before [in red ($+$) and blue ($-$)] and after [orange ($-$)] the interacting region, as indicated by colored dots on the main scheme, with $x=2x_K^+$ and $T_-=10^{-3}T_+$. The coupling $\lambda$ to a probe terminal p can be used to inject thermalized electrons with a temperature similar to the effective temperature after the interaction region.
}
\end{figure}

Here, we suggest the nonlocal thermoelectric response in a multiterminal and multichannel quantum Hall setup to probe the nonequilibrium physics of interacting quantum systems. 
It requires an energy dependent junction connecting terminals at different temperatures~\cite{sanchez_chiral_2015}.
We propose a two-channel configuration where a hot terminal at temperature $T_+$ injects electrons into the inner channel 
while the outer channel connects two other terminals which have identical temperature $T_-<T_+$ with transmission probability ${\cal T}(E)$, see Fig.~\ref{fig:qhsetup}. In this configuration, for all terminals grounded, the nonlocal thermoelectric response of the two separate channels is hence zero, unless they exchange energy via Coulomb interactions~\cite{hotspots,tunneling}. Then a thermocurrent 
\begin{equation}
\label{eq:curr}
I=-\int{dE}\mathcal{T}(E)\left[{f}_{eq}(E,T_-)-{f}_-(x,E)\right]\ ,
\end{equation}
between the conductor terminals with the same temperature $T_-$
pinpoints the presence of e-e interactions in the region of length $x$ (we set $e=\hbar=k_{\rm B}=1$).
The thermocurrent depends on the difference between the nonequilibrium distribution of the outer channel ${f}_-(x,E)$ after the interacting region, and the equilibrium Fermi distribution $f_{eq}(E,T_-)$ of the current terminals 1 and 3~\cite{sivan:1986,butcher:1990}. The longer $x$, the larger the generated current, see Fig.~\ref{fig:qhsetup}(a) for the case of a QPC scatterer, as expected. The deviation of ${f}_-(x,E)$ [plotted in Fig.~\ref{fig:qhsetup}(b)] from an equilibrium distribution can be interpreted as an overheating effect that can be eventually compared with the equilibrium distribution of a probe with an equivalent effective temperature $T_p$.

In particular, we investigate how 1D electronic systems out-of-equilibrium restore thermal equilibrium in the presence of e-e interactions using the nonlocal thermoelectric response as a detector. Other sources of heat transfer (e.g. phonons) can be excluded at sufficiently low temperatures. The analysis is done in the integrable zero-range interactions limit assuming, for simplicity, Gaussian (thermal) current-current correlations in the source contacts. We discuss how the heat transport and noise properties are determined by a universal length scale which depend only on the interaction strength and show differences and similarities in the equilibration processes between a Tomonaga-Luttinger (TL)~\cite{Tomonaga1950,Luttinger1963} and a Landau-Fermi liquid. Furthermore, we use the thermoelectric signature to investigate the crucial role of the cross-correlations in the process of thermal equilibration. 

\emph{Interacting channels.}-- 
We consider copropagating edge states of a quantum Hall fluid at $\nu{=}2$ which interact long enough to reveal the e-e interactions~\cite{fujisawa_review_2022}. Using bosonization~\cite{Haldane1981Jul,vonDelft1998Nov}, 1D ECs can be described in terms of bosonic phase-fields $\phi_\pm(x',t)$. Assuming that zero-range e-e interactions between the two channels ($\alpha,\beta=\pm$) are restricted to a finite region of length $x$, the equations of motion (EoM)
for $0\leq x'\leq x$ become $\partial_t \phi_\alpha +\sum_\beta v_F u_{\alpha\beta} \partial_{x'}\phi_\beta=0$, where $v_F$ is the Fermi velocity. The density-density zero range interaction, $u_{\alpha\beta}=\delta_{\alpha\beta}(1+u_4)+(1-\delta_{\alpha\beta})u_2$, describes intra- ($u_4$) or inter-($u_2$) channel terms~\cite{braggio_1f_2012,fujisawa_review_2022}. Using the current operators at $x=0$, $\hat{J}_\alpha(t)\equiv -\partial_t\phi_\alpha(0,t)/(2\pi)$, as boundary conditions~\cite{levkivskyi_energy_2012,vonDelft1998Nov}, the EoM are analytically  solved by introducing charge and dipole modes $\phi_{c/n}(x,t)= [\phi_+(x,t)\pm \phi_-(x,t)]/\sqrt{2}$, chiral copropagating fields
with velocities $v_{c/n}= v_F (1 + u_4 \pm u_2)$. We assume $u_4=u_2$, so 
$v_n=v_cK=v_F$
with $K\equiv 1/(1+u_4+u_2)$~\cite{acciai_probing_2018}. 
The normal modes chirality,  $\phi_\sigma (x,t)= 
\phi_\sigma (0, t - t_\sigma^x)$ with $t_\sigma^x=x/v_\sigma$ and $\sigma$=c,n connects the phase field operators with the boundary conditions such that the channel current operators $j_\pm(x,\omega)= \frac{1}{2}\sum_{\alpha=\pm}\left(e^{i\omega t_c^x} \pm  \alpha e^{i\omega t_n^x}\right)\hat J_\alpha(\omega)$.
We used the linearity of the theory with the Fourier representation $f(\omega)=\int dt e^{i\omega t}f(t)$. Correlations between the two EC induced by e-e interactions at a given position $x$ can be inspected by looking at the noise spectral density, $S^{\alpha \beta}_x(\omega)=\delta(\omega+\omega')\langle j_\alpha(x,\omega)j_\beta(x,\omega')\rangle$. Note that the $\delta$-function in the r.h.s. is due to the linearity of the theory stating that bosonic modes at different energies are independent quantities. In general, e-e interactions influence both finite auto- $S_{x'}^{\alpha \alpha}(\omega)$ and cross-correlations $S_{x'}^{\alpha \beta}(\omega)$ (with $\alpha\neq \beta$) at any point $x'$ which can be expressed in terms of the boundary correlators $S_0^{\alpha \beta}(\omega)=\langle \hat J_\alpha(\omega)\hat J_\beta(-\omega)\rangle$:
\begin{gather}
\label{eq:sx}
\begin{aligned}
S_x^{\alpha \alpha}(\omega)&=\bar S_0^{\alpha \alpha}(\omega)+\frac{\cos(\omega \delta t^x)}{2}\delta S_0^{\alpha \alpha}(\omega)\\
S_x^{{\alpha} \bar{\alpha}}(\omega)&=-\frac{i}{2}\sin(\omega \delta t^x)\delta S_0^{\alpha \alpha}(\omega),
\end{aligned}
\end{gather}
where $\bar{S}_x^{\alpha \beta}(\omega)\equiv (S_x^{\alpha \beta} + S_x^{\bar\alpha \bar\beta})/2$, $\delta{S}_x^{\alpha \beta}\equiv S_x^{\alpha \beta} - S_x^{\bar\alpha \bar\beta}$,
with $\bar{\alpha}=-\alpha$, and where $\delta t^x=t_n^x-t_c^x= x (1-K)/v_F$ depends on the interaction strength $0 < K \leq 1$. The unsymmetrized (emission) noise spectral density for the noninteracting leads at temperature $T_\alpha$ is $S_0^{\alpha \alpha}(\omega)= \omega [1- e^{-\omega/T_\alpha}]^{-1}e^{-\omega/\omega_c}$, 
with the high-energy cut-off $\omega_c$ ~\cite{levkivskyi_energy_2012}. However,  
finite cross-correlations, $S_x^{{\alpha} \bar{\alpha}}(\omega)$, develop due to e-e interactions, see 
Eq.~\eqref{eq:sx}, which are absent for independent channels. Notably the cross-correlators satisfy the symmetry $S_x^{\alpha\bar{\alpha}}(\omega)=S_x^{\bar{\alpha}\alpha}(\omega)^*$ and are finite only when $S_0^{++}(\omega)\neq S_0^{--}(\omega)$, such as in nonequilibrium for $T_+\neq T_-$. The correlation effects and the oscillating behaviour with $x$ or $\omega$ in Eq.~\eqref{eq:sx} are potentially measurable --and are a signature only of-- a nonequilibrium situation \emph{with} the presence of e-e interactions $K<1$~\cite{suppl}. Further, the period in position (frequency) for fixed frequency (position) is a measure of the interaction strength.  In the following we show two different configurations in which the role of auto- and cross-correlations, and the effects of e-e interactions, can be directly probed by the thermoelectrical response.
\begin{figure}[t!]
\includegraphics[width=\linewidth]{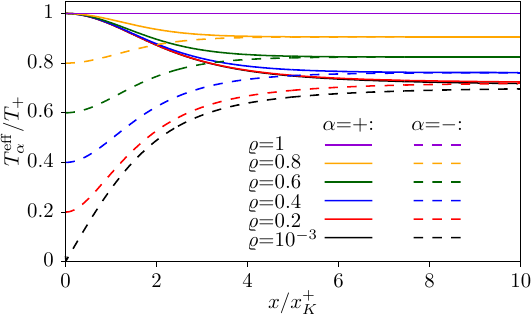}

    \caption{Effective temperature $T^{\rm eff}_\alpha/T_+$ of the $\alpha{=}\pm$ channels vs the rescaled position $x/x_K^+$ for different ${\varrho}=T_-/T_+$. 
}
\label{fig:effTK}
\end{figure}

\emph{Energy exchange and effective temperature}--
When the two channels are out-of-equilibrium, such that $T_+\neq T_-$, it is interesting to investigate how the energy transfer $J_{{\rm E}}$ between the two EC depends on the interaction $K$ and the length $x$ of the interacting zone. 
A direct measurement of the energy exchange is challenging, however it manifests in the EC effective temperature along $x$, which can be probed~\cite{meair_local_2014}.
In particular, we assume that, using QPCs, we can completely separate the ECs after the interaction and analyze them separately~\cite{suppl}. In the limit of a long interacting zone $x\to \infty$, the energy exchanged becomes independent of distance and interaction strength $J_{{\rm E}} \propto T_+^2 - T_-^2$. This is consistent with the expectation that after a long distance, the two interacting ECs equilibrate towards the equilibrium temperature $T^{\rm eq}=\sqrt{T_-^2{+}T_+^2}$ reached by two Fermi liquids with different temperatures put in contact. For an isolated EC for $x'>x$, one can similarly define an effective temperature $T_\pm^{\rm eff}(x')=\sqrt{T_\alpha^2{\mp}\frac{12}{\pi} J_{{\rm E}}(x')}$ which corresponds to the temperature of a Fermi liquid thermalized with the EC~\cite{suppl,integrability}.
For the ECs we find
\be
\label{eq:effTK}
\frac{T_\pm^{{\rm eff}}(x)}{T_+}{=}\sqrt{\frac{1}{2}\mp\frac{3/2}{\sinh^2\tilde x}+\frac{{\varrho}^2}{2}\left[1 \pm \frac{3}{\sinh^2({\varrho} \tilde x)}\right] }, 
\ee
where $\tilde x\equiv x/x_K^+$ is the rescaled interacting length, and ${\varrho}\equiv T_-/T_+$ represents the initial temperature unbalance between the channels. The interaction-dependent effective length $x_K^+ = {\hbar}v_F/[\pi (1{-}K)k_{\rm B}T_+]$ is the typical distance over which the exchange of energy happens between the hot $T_+$ and the cold ($T_-\to 0$) channels.
In Fig.~\ref{fig:effTK} we show the scaling of $T_{\pm}^{{\rm eff}}$ as a function of $\tilde x$ for different ratios ${\varrho}$. Increasing the interaction length $x \gg x_K^+$ the effective temperature equilibrates to the intermediate value of Fermi liquids, $T^{{\rm eq}}/T_+= \sqrt{(1{+}\varrho^2)/2}$. In the limit ${\varrho}\to 0$, Eq.~\eqref{eq:effTK} tends to a universal curve leading to a universal formula for the thermal conductance $G_{th}(x)\approx J_E(x)/\delta T$ assuming initially a small temperature difference $\delta T$ between the ECs~\cite{suppl}.
\begin{figure}[t]
\includegraphics[width=\linewidth]{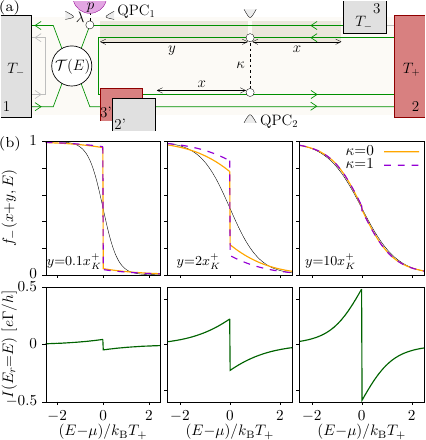}
\caption{(a) Controlled nonequilibrium-  and crosscorrelation-sensitive circuit. Channels at the upper and lower edge are uncoupled for $\kappa{=}0$. Switching on the connection $\kappa{=}1$, inner electrons from terminal 2 are replaced by others from 2'. 
(b) nonequilibrium electron distribution function for the cold EC as a function of energy $E/T_+$ for different distribution functions: fully interacting ($\kappa=0$), and with crosscorrelation resetting ($\kappa=1$), compared with the equilibrium Fermi distribution with temperature $T_-^{{\rm eff}}(y{+}x)$ (black). Different panels refer to different positions $y/x_K^+$, with ${\varrho}{=}10^{-3}$. The lower row shows the current for the correspondingly above cases ($\kappa=0$) when the scatterer is a narrow resonance, $\mathcal{T}(E)\approx\Gamma\delta(E-E_r)$.
}
\label{fig:distribution}
\end{figure}

\emph{Circuit theory for correlators}-- As stated before, the presence of e-e interactions and nonequilibrium can generate finite cross-correlations between the EC. 
In particular it can be shown using the locality (due to zero-range assumption) and linearity of the theory~\cite{suppl} that any correlator $S_{y+x}^{\alpha \beta}(\omega)$ can be expressed in terms of a chain rule via correlators $S_x^{\alpha \beta}(\omega)$ at intermediate position $x$:
\be
\label{eq:chain}
S_{y+x}^{\alpha \beta}(\omega)= \bar{S}_x^{\alpha \beta}+\frac{\cos(\omega\delta t^y )}{2}\delta S_x^{\alpha \beta}+i \frac{\sin(\omega \delta t^y)}{2}\delta S_x^{\alpha \bar{\beta}},
\ee
with the symmetric and anti-symmetric combinations of spectral densities taken at the boundary conditions. 
This result generalizes the second order correlators of Eq.~\eqref{eq:sx} for nonequilibrium situations and potentially nongaussian boundary conditions.
The same chain rule can be used to {\it effectively} describe a region where there is eventually the long range interaction between EC as long as the correlators are investigated in regions where the interaction is again zero-range. Indeed the correlators $S_x^{\alpha\beta}(\omega)$ in a zero-range region contain all the information about averaged quantities (e.g. charge/energy currents) and the noise fluctuations (e.g. second-order current-current correlators) in the system.
This result can be interpreted as a generalized nonequilibrium circuit theory due to the system integrability. One could easily combine regions with different values of interactions. Note in Eq.~\eqref{eq:chain} the influence of cross(auto)-correlators on the later-time auto(cross)-correlators via the term proportional to $i\sin(\omega \delta t^y)$.
This shows that the cross-correlations generated in an interacting system out-of-equilibrium \emph{cannot} be ignored to investigate the energy flow. Below we introduce  an experimental setup able to switch on and off the cross-correlation contributions in a controlled way.

\emph{Nonequilibrium electron distributions.}-- The fundamental quantity in ECs is the electron distribution function ${f}_\alpha (y+x, t)\equiv \langle \psi_\alpha^\dagger (y{+}x ,t)\psi_\alpha(y{+}x,0)\rangle$ where, in bosonization, the operators are expressed in terms of phase fields since $\psi_\alpha (x,t)\propto e^{-i \phi_\alpha(x,t)}$.  
However, in order to address the role of cross-correlations in the energy exchange mediated by the interaction we investigate the thermoelectricity in a specific setup, see Fig.~\ref{fig:distribution}(a). We consider two EC separately contacted with ohmic contacts at different temperatures $T_\alpha$ at $x<0$. They start to interact from $x\geq 0$, and we measure the thermoelectrical response induced by the outer EC at the scatterer $\mathcal{T}(E)$ located at $y+x>0$. 
At an intermediate position $x$ one can imagine to selectively open, $\kappa=1$, (close, $\kappa=0$) the QPC$_2$, switching on (off) the cross-correlation~\cite{cross}.

The real-time distribution function is thus given by ${f}_\alpha (y+x,t)\propto e^{i g_\alpha (y+x,t)}$ with the exponent~\cite{suppl}
\be
\label{eq:exponent}
g_\alpha(y+x,t)= \int_{-\infty}^{+\infty}d\omega \big[e^{-i\omega t}-1\big]\frac{S_{y+x}^{\alpha \alpha}(\omega)}{\omega^2}~.
\ee
The energy-dependent distribution function of the cold EC ${f}_-(y+x,E)$ (the Fourier transform of the above expressions)
is plotted in the top panels of Fig.~\ref{fig:distribution}(b) in the limiting case where it was prepared at almost zero temperature ${\varrho}\to 0$. 
Different curves refer to different distribution functions: red for the full-interacting case, blue in absence of cross-correlations at $x$, i.e. $S_{x}^{{\alpha} \bar{\alpha}}(\omega)=0$. As a reference we have also plotted the Fermi distribution function evaluated with the same effective temperature $T_-^{{\rm eff}}(y+x)$, see black solid line in Fig.~\ref{fig:distribution}(b).
One immediately notes the nontrivial role of cross-correlations of the nonequilibrium distribution functions. 
The electron distribution differences are mainly due to the power-law behaviour around $\mu=0$ typical of a TL liquid. Importantly, the effects of cross-correlations on electron distributions are mainly relevant at intermediate distances $y\sim x_K^+$. At short distances the term $i\sin[\omega x(1-K)/v_F]$ suppresses their contribution. 
For long distances $y\gg x_K^+$, ${f}_-(y+x,E)$ with and without cross-correlations become equal, but still slightly different from the equilibrium Fermi distribution function. This is a consequence of the constraints dictated by integrability. However, the system spontaneously seems to develop a Fermi-like behaviour even if it slightly differs from Fermi distribution at $\mu$, a clear hallmark of the e-e interaction~\cite{altimiras_nonequilibrium_2010,lesueur_energy_2010,itoh_signatures_2018}.

\emph{Thermoelectric currents.}-- 
Eq.~\eqref{eq:curr} gives a direct way to detect the deviation of the cold EC ($-$) from the equilibrium distribution via the thermoelectric current. For this, we use a narrow-band spectrometer~\cite{altimiras_nonequilibrium_2010,lesueur_energy_2010} such as a resonant antidot~\cite{buttiker_negative_1988} with transmission $\mathcal{T}(E)\approx\Gamma\delta(E{-}E_r)$, which has the dual property of being an efficient thermoelectric~\cite{josefsson_quantum_2018} as well as being sensitive to the features near the tunable resonant energy $E_r$, i.e. $I=-e\Gamma\left[{f}_{eq}(E_r,T_-)-{f}_-(x,E_r)\right]$, see bottom panels of Fig.~\ref{fig:distribution}(b). As expected the current increases with the growing of the interaction distance $y{+}x$, since more energy is transferred from hot to the cold EC. Clearly, the thermocurrent saturates when $y\gg x_K^+$ (not shown).

The setup of Fig.~\ref{fig:distribution}(a) provides the unique possibility to test in a DC measurement the difference in the energy exchange between the interacting and the noninteracting case or, even, how the suppression of cross-correlations influence the energy transfer. 
The interaction effects are measured by tuning QPC$_1$, which couples the outer channel to a probe terminal with a transmission probability $\lambda$. 
This way we can directly compare the thermoelectric current generated from the noninteracting Fermi distribution of the probe ($\lambda{=}1$) with the distribution coming from the interaction region ($\lambda{=}0$).
This is represented by the difference of thermocurrents $\delta I_\lambda=I(\lambda{=}0)-I(\lambda{=}1)$. The comparison is particularly meaningful as the probe has the same temperature of the cold EC channel entering, i.e. $T_+^{{\rm eff}}(x+y)$~\cite{meair_local_2014}. 

The effect of cross-correlations can be, instead, detected by acting on QPC$_2$, which couples the upper and lower inner channels with transition probability $\kappa$. In this case it is important to note that there is a copy of terminals 2 and 3 that we call 2' and 3', which symmetrically operate on the ECs of the other side of the Hall bar.
For $\kappa=0$, the two upper channels interact along a distance $x+y$ as discussed above. 
Switching on the connection, $\kappa=1$, one replaces the inner channel with another one having nominally the same auto- but no cross-correlations at $x$ (as they did not have the possibility to interact before).
The effect of cross-correlations in the current in hence is represented by the difference $\delta I_\kappa=I(\kappa{=}0)-I(\kappa{=}1)$. 
\begin{figure}[t]
\includegraphics[width=\linewidth]{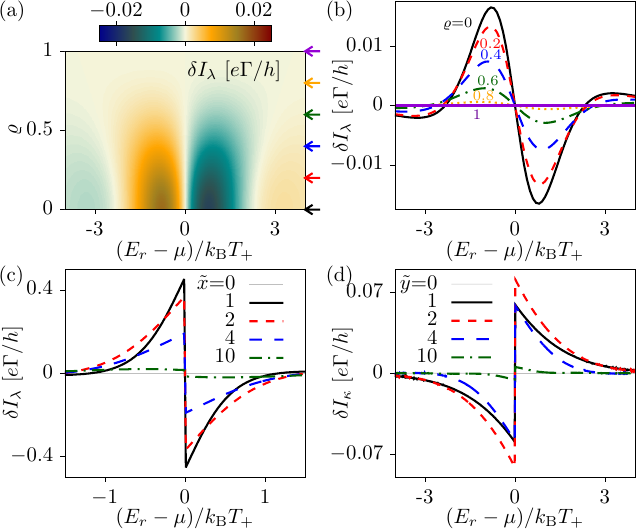}
\caption{\label{fig:frog} (a) Difference of the currents with and without the probe, $\delta I_\lambda$, for $T_+{=}1$ and $\tilde x{=}300$, corresponding to an antidot-like scatterer with $\mathcal{T}(E)=\Gamma\delta(E{-}E_r)$, as a function of the resonance energy and $T_-$. (b) Cuts of the previous for fixed values of $T_-$ marked by the corresponding color arrows in panel (a). (c) $\delta I_\lambda$ for different values of $\tilde x$ and ${\varrho}=10^{-3}$. (d) $\delta I_\kappa$ when resetting cross-correlations at a distance $\tilde x$ and before another distance $\tilde y=y/x_K^+$, for ${\varrho}=10^{-3}$ and $\lambda=0$. }
\end{figure}

\emph{Interaction and correlation effects}-- We firstly consider to control $\lambda$ of the QPC$_1$ keeping fixed the QPC$_2$ ($\kappa=0$).  
This is shown in Figs.~\ref{fig:frog}(a) and (b) for different initial reservoir temperatures $T_-$ varying the resonance energy $E_r$. 
A finite $\delta I_\lambda$ means that the electronic distribution emerging from the interacting zone is different from a Fermi distribution. Thermoelectricity is sensitive to this difference, especially in the energy window around $\mu$. Obviously, the thermocurrent generally grows with the increasing temperature difference.
Note that $\delta I_\lambda$ is an odd function around $\mu=0$ and it is maximal when $|E_r-\mu|\sim k_{\rm B}T_+$, roughly the energy scale where the two-electron distributions are maximally different~\cite{narrow}.  However, for the long interaction length limit $y+x\gg x_K^+$, the interacting electron distribution becomes Fermi-like and $\delta I_\lambda$ gets reduced, see Figs.~\ref{fig:frog}(c). 

If instead we control $\kappa$ of the QPC$_2$ keeping QPC$_1$ fixed at $\lambda=0$, we can observe some intriguing effects of cross-correlations. 
In Fig.~\ref{fig:frog}(d) we consider the case $x=x_K^+$ for different distances $y$, finding a nonmonotonous behavior in $y$. For $y\ll x_K^+$ the cross-correlations do not have enough space to develop an influence on the autocorrelations due to the sine prefactor of Eq.~\eqref{eq:chain}, so we expect $\delta I_\kappa\approx 0$.  
At the same time for $y\gg x_K^+$ we also expect $\delta I_\kappa\to 0$ because the resulting interacting electron distribution is Fermi-like [see also Fig.~\ref{fig:distribution}(b)] and the cross-correlation does not play any major role in this limit. The role of the correlations in the energy flow between the two channels is maximal for $y\approx x_K^+$ and for $|E_r-\mu|\sim k_{\rm B}T_+$, see Fig.~\ref{fig:frog}(d). 

\emph{Conclusions.}-- We propose the nonlocal thermoelectric response as a signature of the presence of interacting nonequilibrium states in copropagating quantum Hall channels. We identify the characteristic length over which energy is mainly exchanged due to the interaction, which is manifested in the generated thermocurrent and gives a direct measure of the interaction strength. Controlling the connections between the different edge channels, the importance of cross-correlations in the energy flow of interacting nonequilibrium system is quantified in a purely DC setup, opening ways to applications of thermoelectricity in quantum sensing. Our results introduce a circuit theory that can be extended to configurations with other kinds of edge states (e.g. in topological insulators), or with long-range interactions with proper modifications.

\begin{acknowledgments}
A.B. acknowledges the MIUR-PRIN2022 Project NEThEQS (Grant No. 2022B9P8LN), the Royal Society through the International Exchanges between the UK and Italy (Grants No. IEC R2 192166) and EU’s Horizon 2020 Research and Innovation Framework Programme under Grant No. 964398 (SUPERGATE) and No. 101057977 (SPECTRUM). M.C. acknowledges support from PRIN MUR Grant No. 2022PH852L. B.S. acknowledges financial support from the Ministry of Innovation NRW via the ``Programm zur Förderung der Rückkehr des hochqualifizierten Forschungsnachwuchses aus dem Ausland'' and Deutsche Forschungsgemeinschaft (DFG, German Research Foundation) - Project-ID 278162697 – SFB 1242. R.S. acknowledges funding from the Ram\'on y Cajal program RYC-2016-20778, and the Spanish Ministerio de Ciencia e Innovaci\'on via grant No. PID2019-110125GB-I00 and through the ``Mar\'{i}a de Maeztu'' Programme for Units of Excellence in R{\&}D CEX2018-000805-M, and the COST Action MP1209 short term scientific mission during which this project was initiated. 
\end{acknowledgments}

\bibliography{biblio}

\begin{thebibliography}{61}
\bibitem [{tun()}]{tunneling}%
  \BibitemOpen
  \href@noop {} {}\bibinfo {note} {{We are neglecting direct electron
  tunnelling between the two channels since we assume the equilibration length
  between the two channels longer than the sample size.}}\BibitemShut {Stop}%
\bibitem [{int()}]{integrability}%
  \BibitemOpen
  \href@noop {} {}\bibinfo {note} {{However, due to the integrability, the
  electronic distribution of the EC can be also quite different from a Fermi
  distribution so that effective temperature is a rough estimation of thermal
  energy flow in the channel.}}\BibitemShut {Stop}%
\bibitem [{cro()}]{cross}%
  \BibitemOpen
  \href@noop {} {}\bibinfo {note} {{The crosscorrelations emerge due to the
  interaction along the distance $x$ for an open QPC$_2$ ($\kappa=1$) but, when
  closed ($\kappa=0$) the two ECs comes from different circuits, which avoids
  cross-correlations.}}\BibitemShut {Stop}%
\bibitem [{nar()}]{narrow}%
  \BibitemOpen
  \href@noop {} {}\bibinfo {note} {{This means that the narrow-resonance
  assumption can be even relaxed up to $\Gamma\sim k_{\rm B}T_+$ resulting in
  larger thermocurrent~\cite{suppl}.}}\BibitemShut {Stop}%
\end{thebibliography}
\pagebreak
\widetext
\clearpage


\section*{Supplementary material}

# Supplementary Material for "Nonlocal thermoelectric detection of interaction and correlations in edge states"

Alessandro Braggio,[1] Matteo Carrega,[2] Björn Sothmann,[3] and Rafael Sánchez[4]

[1]*NEST, Istituto Nanoscienze-CNR and Scuola Normale Superiore, Piazza San Silvestro 12, I-56127 Pisa, Italy*
[2]*SPIN-CNR, Via Dodecaneso 33, 16146 Genova, Italy*
[3]*Theoretische Physik, Universität Duisburg-Essen and CENIDE, D-47048 Duisburg, Germany*
[4]*Departamento de Física Teórica de la Materia Condensada, Condensed Matter Physics Center (IFIMAC),
and Instituto Nicolás Cabrera, Universidad Autónoma de Madrid, 28049 Madrid, Spain*
(Dated: July 18, 2023)

## NOISE SPECTRAL DENSITY AT POSITION $x$

We calculate the current noise spectral density at the position $x$ as a function of the noise spectral density injected at the initial position $x = 0$. The noise spectral density can be generally defined in terms of the current-current correlation function

$$\langle j_\alpha(x,\omega) j_\beta(x,\omega') \rangle = \delta(\omega + \omega') \langle j_\alpha(x,\omega) j_\beta(x,-\omega) \rangle = \delta(\omega + \omega') S_x^{\alpha\beta}(\omega), \tag{S1}$$

where the first equality is directly connected to the fact that charge plasmonic modes at different energies are independent as a consequence of the linearity of the theory. Hereafter we also assume that the spectral densities at the initial boundary $x = 0$ of the two channels are completely uncorrelated i.e.

$$\langle j_\alpha(x=0,\omega) j_\beta(x=0,\omega') \rangle = \delta(\omega + \omega') \langle J_\alpha(\omega) J_\alpha(-\omega) \rangle = \delta_{\alpha\beta} \delta(\omega + \omega') S_0^{\alpha\alpha}(\omega) \ , \tag{S2}$$

where $S_0^{\alpha\alpha}(\omega)$ is the boundary spectral density. For non-interacting leads one easily finds

$$S_0^{\alpha\alpha}(\omega) = \frac{\omega}{1 - e^{-\omega/T_\alpha}} e^{-|\omega|/\omega_c} \tag{S3}$$

where $T_\alpha$ is the temperature ($k_B = \hbar = 1$) of the lead connected to the mode $\alpha = \pm$. We also include the usual cut-off energy scale $\omega_c$, of the order of Fermi energy, choosing for it the exponential form. We observe that the previous form has two important limits: high temperatures ($\omega_c \gg T \gg \omega$), where $S_0^{\alpha\alpha}(\omega) \approx T$ with dominating thermal fluctuations, and low temperatures ($\omega_c \gg \omega \gg T$), where the spectrum is dominated by quantum fluctuations and $S_0^{\alpha\alpha}(\omega) \approx \omega$.

The expression of the current operators in frequency space are

$$j_\pm(x,\omega) = \frac{1}{2} \left[ \left( e^{i\omega t_c^x} + e^{i\omega t_n^x} \right) J_\pm(\omega) + \left( e^{i\omega t_c^x} - e^{i\omega t_n^x} \right) J_\mp(\omega) \right] \tag{S4}$$

where $t_\sigma^x$ is introduced in the main text and we wrote the current operator in terms of charged and neutral eigenmode time evolution. Substituting those operators into Eq. (S1), one finds that the autocorrelated spectral function $S_x^{\alpha\alpha}(\omega)$ at the position $0 \le x \le L$ for the two modes depends only on the spectral functions at the boundaries, i.e. $S_0^{\alpha\alpha}(\omega)$, indeed

$$S_x^{\alpha\alpha}(\omega) = \frac{1 + \cos[\omega(t_n^x - t_c^x)]}{2} S_0^{\alpha\alpha}(\omega) + \frac{1 - \cos[\omega(t_n^x - t_c^x)]}{2} S_0^{\bar\alpha\bar\alpha}(\omega) \tag{S5}$$

where $\bar\alpha = -\alpha$. Firstly we observe that the contribution of the two modes (the same or the opposite mode) has a different sign in front of the cosine term. This result is quite general and can be interpreted as the consequence of the difference in the propagation of the neutral and charged modes as obtained for the zero-range interaction between the two channels along a distance $x$.

Assume now that the two channels are kept initially at different temperatures as considered in the setup discussed in main text. Considering the limiting case $T_+ > T_- \gg \omega$, and using that $t_n^x - t_c^x = x(1-K)/v_F$, we get

$$S_x^{\pm\pm}(\omega) \approx \frac{T_+ + T_-}{2} \pm \frac{T_+ - T_-}{2} \cos[\omega x(1-K)/v_F] \tag{S6}$$

The above limit shows that at high temperatures and for short distances $x \ll \pi v_F/[(1-K)\omega]$ the two modes will start to equilibrate toward and intermediate averaged temperature. However, this equilibration is only apparent and

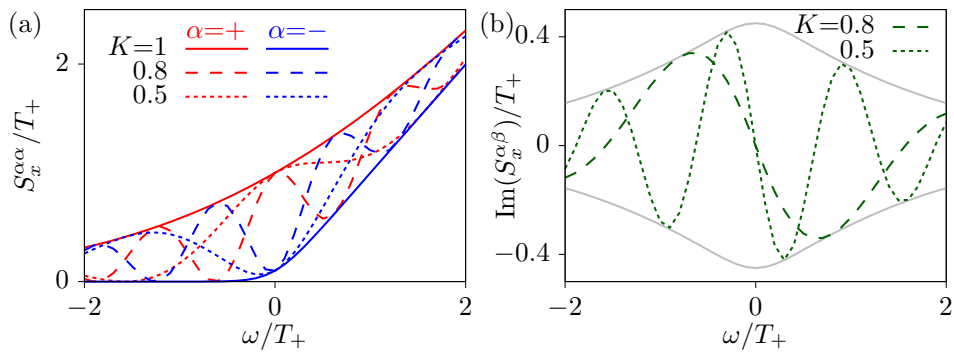

FIG. S1. (a) Rescaled autocorrelation spectral densities $S_x^{\alpha\alpha}/T_+$ ($\alpha=+$, in red, $\alpha=-$, in blue lines) calculated in the position $x$ such that $xk_BT_+/(\hbar v_F) = 10$ as a function of the rescaled frequency $\omega/T_+$ for $T_+ = 10T_-$. The different linestyles correspond to different values of the interaction parameter, as indicated. The solid lines practically represent the initial spectral densities at the initial temperatures and we see that with interaction the spectral density oscillate between those two limiting spectra. (b) Rescaled imaginary part of the cross-correlation spectral densities $\mathrm{Im}[S_x^{\alpha\beta}]/T_+$ with $\beta = -\alpha$ calculated in the position $x$ such that $xk_BT_+/(\hbar v_F) = 10$ as functions of the rescaled frequency $\omega/T_+$. The solid grey curves represent respectively the difference of initial spectral density $\pm(S_0^{\alpha\alpha} - S_0^{\bar\alpha\bar\alpha})/2$ which are the envelope for the oscillating behaviour, with the different linestyles corresponding to different interaction parameters as before.

appears, for a fixed energy $\omega$, only at specific intermediate points. The cosine term describes an oscillating behaviour of the energy exchange between the two channels, which indeed depends on the interaction length $x$ and the energy $\omega$. This sort of oscillatory plasmonic effect in the energy exchange may bring to the conclusion that an initially hotter channel could eventually get lower correlations (as corresponding to a lower temperature) and viceversa. However, this is not the case as we will see, since a certain temperature state is an incoherent superposition of different plasmonic modes at different energies. In conclusion from the full superposition of all the energy channels, the thermal energy is gradually exchanged between the two modes loosing the oscillatory component, as described in Fig. 2 of the main text. This of course agrees with the second principle of thermodynamics which prescribes that thermal energy naturally flows only from hot to cold.

In Fig. S1(a) we show the non-symmetrized spectral densities as functions of frequency at a fixed distance $x$ and for different values of the interaction parameter. We firstly note that for the non-interacting case with $K = 1$ charge and dipole modes travel at the same speed, implying no oscillatory behaviour, so $S_x^{\alpha\alpha}(\omega) \equiv S_0^{\alpha\alpha}(\omega)$ at any $x$. Increasing the interaction the oscillatory behaviour becomes visible. Note that the period of the oscillations is an indirect measure of the interaction parameter $K$. At high frequencies $\omega \gg T_\pm$ the quantum regime emerges making the spectra of the two modes identical and featureless, $S_x^{\alpha\pm}(\omega) \approx \omega$, as one would expect in the quantum noise regime.

Following the same steps we can calculate the cross-correlation:

$$S_x^{\alpha\bar\alpha}(\omega) = -i\frac{\sin[\omega(t_n^x - t_c^x)]}{2}\left(S_0^{\alpha\alpha}(\omega) - S_0^{\bar\alpha\bar\alpha}(\omega)\right) = [S_x^{\bar\alpha\alpha}(\omega)]^*, \tag{S7}$$

where the last passage reflects the obvious relation between crosscorrelators. We can see some similarity with the structure of the autocorrelations $S_x^{\alpha\alpha}$ but also striking differences. In particular we see the cross-correlator is purely imaginary, at least assuming that source correlations does not contain any imaginary component. This shows that the two current operators differ exactly of a $\pi/2$ phase. Note that the multiplicative presence of the sine imposes that the cross-correlation may change sign, see Fig. S1(b). By the fact that the cross-correlators are purely imaginary and are linear combination of the initial spectral densities, and from the relation $S_0^{\alpha\alpha}(\omega) = [S_x^{\bar\alpha\alpha}(\omega)]^*$, one immediately sees that it must be proportional to the difference $S_0^{\alpha\alpha}(\omega) - S_0^{\bar\alpha\bar\alpha}(\omega)$. As a consequence, no cross-correlations develop when the two channels have initially the same temperature, i.e. they are a manifestation of non-equilibrium effects mediated by interactions. The phase of the oscillation directly depends on the sign of $(1 - K)$, which changes sign between repulsive ($K < 1$) and attractive regime ($K > 1$).

## THERMAL CURRENTS

Let us focus on the energy current in channel $\alpha$, which contains information on the mutual influence between the two modes. We consider the energy current after the interacting region of length $x$ in the absence of any bias difference

applied between the two channels. One can use the locality of the theory (due to zero-range interaction) to properly define the total energy current operator $J_\mathrm{E}^\alpha$ at any point $x$, including in the interacting zone. This can be done by firstly observing that the total energy flow along the edge is given by the sum of the Hamiltonian density of the normal modes $\mathcal{H}_\sigma$:

$$\sum_{\sigma=c,n} v_\sigma \mathcal{H}_\sigma(x,t) = \sum_\alpha J_\mathrm{E}^\alpha(x,t) \tag{S8}$$

where $v_\sigma$ are the normal mode velocities. In the r.h.s we report that the energy current operator for the $\alpha$ modes clearly divide the contribution in terms of the contribution of each channel to the total flow. To define a physically consistent operator for $J_\mathrm{E}^\alpha(x)$ we need to use the chirality in the problem. We assume that at the point $x$ we physically separate the two channels, such that, just at the measurement point $x$, they do not interact anymore. So we can calculate the energy current just after the interacting zone (at $x+0^+$) where the energy operator is simply given by the non-interacting expression being

$$J_\mathrm{E}^\alpha(x,t) = \frac{v_\mathrm{F}^2}{4\pi}\left(\partial_x\phi_\alpha(x+0^+,t)\right)^2 = \frac{1}{4\pi}\left(\partial_t\phi_\alpha(x+0^+,t)\right)^2 = \pi j_\alpha(x,t)^2. \tag{S9}$$

In the first identity we write the energy current as the non-interacting Hamiltonian density multiplied by the free propagation velocity $v_F$. In the second identity we used the chirality condition for the free propagation and the last one comes directly from the definition of the particle current $j_\alpha(x,t) = -\partial_t\phi_\alpha(x,t)/(2\pi)$. This expression is generally valid, for a locally interacting chiral theory, even when $x$ is inside the interacting zone. Indeed this operator prescription cannot change if, for $x' > x$, the system is again short-range interacting, due to the mentioned chirality which implies that operator at $x$ cannot be dependent from operators defined at a later time (position).

Now we express the energy current operators $j_\alpha(x,t)$ in terms of their Fourier components and write

$$\begin{aligned}J_\mathrm{E}^\alpha(x,t) = \frac{1}{16\pi}\int_{-\infty}^{+\infty} d\omega\, d\omega'&\left[\left(e^{i\omega(t-t_c^x)}+e^{i\omega(t-t_n^x)}\right)J_\alpha(\omega)+\left(e^{i\omega(t-t_c^x)}-e^{i\omega(t-t_n^x)}\right)J_{\bar\alpha}(\omega)\right]\\
&\times\left[\left(e^{i\omega'(t-t_c^x)}+e^{i\omega'(t-t_n^x)}\right)J_\alpha(\omega')+\left(e^{i\omega'(t-t_c^x)}-e^{i\omega'(t-t_n^x)}\right)J_{\bar\alpha}(\omega')\right],\end{aligned} \tag{S10}$$

which has to be interpreted as an operatorial identity, with the operators $J_\alpha$ describing the boundary conditions. So the mean energy current is determined by the boundary conditions correlators of Eq. (S2), giving:

$$\langle J_\mathrm{E}^\alpha(x)\rangle = \frac{1}{8\pi}\int_{-\infty}^{+\infty} d\omega\left\{\left(1+\cos[\omega(t_c^x-t_n^x)]\right)S_0^{\alpha\alpha}(\omega)+\left(1-\cos[\omega(t_c^x-t_n^x)]\right)S_0^{\bar\alpha\bar\alpha}(\omega)\right\} = \frac{1}{4\pi}\int_{-\infty}^{+\infty}d\omega\, S_x^{\alpha\alpha}(\omega), \tag{S11}$$

which is time independent, as expected. In order to identify the underlining scattering form of the plasmon modes, it is convenient to express the boundary spectral density

$$S_0^{\alpha\alpha}(\omega) = \omega e^{-|\omega|/\omega_c}[n_B^\alpha(\omega)+1] \tag{S12}$$

in terms of plasmon bosonic population $n_B^\alpha(\omega) = (e^{\omega/T_\alpha}-1)^{-1}$. Thus

$$\langle J_\mathrm{E}^\alpha(x)\rangle = \frac{1}{4\pi}\sum_{\sigma=\pm}\int_0^{+\infty}d\omega\,\omega e^{-\omega/\omega_c}\left(1+\sigma\cos[\omega(t_n^x-t_c^x)]\right)n_B^{\sigma\alpha}(\omega), \tag{S13}$$

where we used the symmetries of the integrand to reduce the frequency integration only over positive frequencies using the relation $n_B^\alpha(-\omega) = -1-n_B^\alpha(\omega)$. This result may be physically interpreted as the contribution to the energy current of channel $\alpha$ as induced by the magnetoplasmons of energy $\omega$ originally injected in the two channels. In particular one could recognise in the previous formula the plasmon population $n_B^\alpha(\omega)$ as injected at $x=0$ in lead $\alpha$ and the prefactors containing the cosine as the scattering matrix contribution calculated to propagate plasmons to the point $x$. We note that in this expression the $+1$ terms summed to the population of (S12), is suppressed since it describes the contribution of the vacuum fluctuations. These terms will diverge with the cutoff. Those vacuum terms are cancelled using the usual normal ordering prescription for the operators where the unmeasurable vacuum contribution are necessarily subtracted.

We are interested in the energy current transferred from the "hot" to the "cold" channel as a measure of the energy transfer mediated by interactions. This corresponds to the "energy loss" in the interacting zone from the hot ($+$) channel ($T_+ > T_-$). We define it as

$$J_\mathrm{E}(x) \equiv \langle J_\mathrm{E}^+(0)\rangle - \langle J_\mathrm{E}^+(x)\rangle. \tag{S14}$$

That is in general $J_{\mathrm{E}}(x) > 0$. This quantity is thus given by

$$J_{\mathrm{E}}(x) = \frac{1}{4\pi} \int_0^{+\infty} d\omega\, \omega \left(1 - \cos[\omega(t_c^x - t_n^x)]\right) \left[n_{\mathrm{B}}^+(\omega) - n_{\mathrm{B}}^-(\omega)\right], \tag{S15}$$

where we have taken also the scaling limit $\omega_c \to \infty$. We observe that in the non-interacting limit, where $t_c^x = t_n^x$, we recover $J_{\mathrm{E}}(x) = 0$. The previous equation can be solved analytically [S1], giving

$$J_{\mathrm{E}}(x) = \frac{\pi}{8} \sum_{\sigma=\pm} \sigma T_\sigma^2 \left(\frac{1}{3} + \frac{1}{\sinh[\pi T_\sigma (t_n^x - t_c^x)]^2}\right) = \frac{\pi}{8} \sum_{\sigma=\pm} \sigma T_\sigma^2 \left(\frac{1}{3} + \frac{1}{\sinh[\pi T_\sigma x(1-K)/v_F]^2}\right), \tag{S16}$$

where in the second identity we explicitly showed the dependence on the interaction parameter $K$, position $x$ and Fermi velocity $v_F$.

At low temperatures (or short distances) $T_\alpha(t_n^x - t_c^x) \ll 1$ the previous expression becomes

$$J_{\mathrm{E}}(x) = \frac{\pi^3}{120}(t_n^x - t_c^x)^2 \sum_{\sigma=\pm} \sigma T_\sigma^4 + \mathcal{O}(T^5) = \frac{\pi^3}{120} \frac{x^2(1-K)^2}{v_c^2} \sum_{\sigma=\pm} \sigma T_\sigma^4 + \mathcal{O}(T_\sigma^5), \tag{S17}$$

where the energy exchange between the modes grows nonlinearly but quadratically as a function of $x$ at short distances, and as $(1-K)^2$, with the interaction.

At high temperatures (or sufficiently long distances $x$) $T_\alpha(t_n^x - t_c^x) \gg 1$ we have

$$J_{\mathrm{E}}(x) = \frac{\pi}{24} \sum_{\sigma=\pm} \sigma T_\sigma^2 \left(1 - 12 e^{-2\pi T_\sigma(t_n^x - t_c^x)}\right) + \mathcal{O}(1/T_\sigma^5), \tag{S18}$$

which in the limit $x \to \infty$ becomes

$$J_{\mathrm{E},\infty} = \frac{\pi}{24} \frac{k_B^2}{\hbar} (T_+^2 - T_-^2) + \mathcal{O}(1/T_\sigma^5). \tag{S19}$$

Here we have explicitly reintroduced the physical constants to provide the reader a better clue on the physical content of the result. Note that in the long distance limit the energy loss of the mode is both independent of distance and of the interaction strength $K$. This is expected, since after a certain distance the two channels equilibrate at an intermediate temperature. From the energy loss one can predict the expected temperature using the following argument. Firstly one should assume that after the point $x$ the two modes do not interact anymore. Thus we can treat them as standard chiral Fermi liquids (CFLs) with Fermi velocity $v_F$. For a CFL the expected particle-hole energy current is given by

$$J_{\mathrm{CFL}}(T) = v_F\, \mathcal{U}_{\mathrm{CFL}} = \frac{\pi}{12} \frac{(k_B T)^2}{\hbar}, \tag{S20}$$

where $\mathcal{U}_{\mathrm{CFL}} = (\pi/12)(k_B T)^2/(\hbar v_F)$ is the energy density of 1D CFL at temperature $T$. So in the non-interacting chiral lead at (zero bias) one could directly associate an effective temperature to a flux of energy using previous formula. Observing that indeed for $x = 0$ and after the position $x$ the system is non-interacting we could define an *effective temperature*, in the position $x$, $T_\alpha^{\mathrm{eff}}(x)$ as

$$T_\alpha^{\mathrm{eff}}(x) = \sqrt{T_\alpha^2 - \alpha \frac{12}{\pi} J_{\mathrm{E}}(x)}. \tag{S21}$$

This way, using the result of Eq. (S16) we find for $T_+ > T_- \geq 0$

$$\frac{T_+^{\mathrm{eff}}(x)}{T_+} = \sqrt{1 - \frac{1}{2}\left(1 - \varrho^2 + \frac{3}{\sinh[x/x_K^+]^2} - \frac{3\varrho^2}{\sinh[\varrho(x/x_K^+)]^2}\right)}, \tag{S22}$$

expressed as a function of the initial temperature ratio $\varrho = T_-/T_+$ and the temperature dependent effective length

$$x_K^\alpha = \frac{v_F}{\pi(1-K)T_\alpha}. \tag{S23}$$

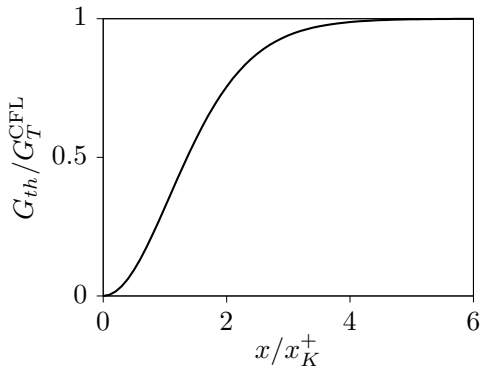

FIG. S2. Universal behaviour of the quantity $G_{th}$ of Eq. (S25), normalized with respect to the quantum of thermal conductance $G_T$, as a function of $x/x_K^+$.

The rescaled effective temperature $T_+^{\text{eff}}/T_+$ of mode $+$, is plotted as a function of the position $x/x_K^+$ in the main text. Note that the equilibration curve in the limit $\varrho \to 0$ becomes universal:

$$\frac{T_+^{\text{eff}}(x)}{T_+} = \sqrt{\frac{1}{2} + \frac{3}{2}\left[\left(\frac{x_K^+}{x}\right)^2 - \frac{1}{\sinh[x/x_K^+]}\right]}. \tag{S24}$$

It only depends on the scaled distance $x/x_K^+$. This curve practically coincides with the solid black curve of the Fig. 2 in the main text.

A similar scaling behaviour is obtained if we linearize Eq. (S16) around an average temperature $\bar{T} = (T_+ + T_-)/2$, assuming a small temperature difference between the two channels $\delta T = T_+ - T_- \ll \bar{T}$. In such regime one finds $J_{\text{E}} \approx G_{th}\delta T$ where

$$G_{th}(x) = \frac{\pi}{4}\bar{T}\left[\frac{1}{3} + \frac{1}{\sinh[x/x_K]^2}\left(1 - \left(\frac{x}{x_K}\right)\coth[x/x_K]\right)\right]. \tag{S25}$$

which is also an universal function of the ratio $x/x_K$, with $x_K \equiv v_F/\pi(1-K)\bar{T}$. This is shown in Fig. S2, rescaled with the thermal conductance of a CFL, $G_T^{\text{CFL}} = \pi\bar{T}/12$ (recall that $k_B = \hbar = 1$). At infinite $x$ we recover $G_{th}(x \to \infty) = -G_T^{\text{CFL}}$ so the effectiveness of energy transfer between the two channels reaches the ideal value of the thermal conductance quantum.

## ELECTRON DISTRIBUTION FUNCTIONS

We now analyse the electron distribution functions assuming simple Gaussian (thermal) correlations in the boundary conditions, such as in the case of standard Ohmic contacts. Using the bosonization formalism, the electron distribution function can be written as

$$\langle \psi_\alpha^\dagger(x,t)\psi_\beta(x,0)\rangle = \frac{1}{2\pi\delta}\langle e^{i\phi_\alpha(x,t)}e^{-i\phi_\beta(x,0)}\rangle \tag{S26}$$

with $\alpha, \beta = \pm$, where $\delta$ is the usual short-distance cut-off in the bosonization language. In the r.h.s. of the above equation $\phi_\pm(x,t)$ are given by

$$\phi_\pm(x,t) = \pi\left[Q_+(t - t_c^x) + Q_-(t - t_c^x) \pm Q_+(t - t_n^x) \mp Q_-(t - t_n^x)\right]. \tag{S27}$$

We assume that at position $x = 0$ the two boundary charge operators $Q_\pm(t)$ are uncorrelated modes with Gaussian distributions. For our purposes here we only need the electronic distribution functions $\langle \psi_\pm^\dagger(x,t)\psi_\pm(x,0)\rangle$:

$$\begin{aligned}
f_\pm(x,t) &\equiv (2\pi\delta)\langle \psi_\pm^\dagger(x,t)\psi_\pm(x,0)\rangle \\
&= \langle e^{i\pi\left(Q_\pm(t-t_c^x) + Q_\pm(t-t_n^x) + Q_\mp(t-t_c^x) - Q_\mp(t-t_n^x)\right)} e^{-i\pi\left(Q_\pm(-t_c^x) + Q_\pm(-t_n^x) + Q_\mp(-t_c^x) - Q_\mp(-t_n^x)\right)}\rangle.
\end{aligned} \tag{S28}$$

To solve this, we will use the identity $\langle e^{Q_1} e^{Q_2} ... e^{Q_n}\rangle = e^{\sum_j \langle Q_j^2/2\rangle} e^{\sum_{i<j}\langle Q_i Q_j\rangle}$ valid in bosonization for thermal averages when the commutator $[Q_i, Q_j]$ is a complex number [S2]. At the same time the charge operators can be decomposed in Fourier components

$$Q_\pm(t) = \int_{-\infty}^{+\infty}\frac{d\omega}{2\pi}e^{-i\omega t}\frac{J_\pm(\omega)}{-i\omega}, \tag{S29}$$

in terms of the boundary current operators $J_\alpha(\omega)$. However, in the following we are interested in specifying the electron distribution at $y + x$ in terms of the current correlators at $x$. These correlations will be finite and given by Eq. (2) of the main text, where it is assumed that the channels have interacted before for a distance $x$. In the main text we consider also the case where the cross-correlation will be suppressed. This describes the case where the two modes come from different terminals, so they never developed cross-correlations even if the auto-correlation are identically prepared to the fully interacting case considered before.

Using the linearity of the theory, we can express the previous charge operator $Q_\pm$ in terms of the operator $j_\alpha(x,\omega)$ by simply inverting the linear equation (S4). The electronic distribution is obtained making the averages of the two point correlators we got finding the cross-correlation $S_x^{\alpha\beta}(\omega)$ as defined in Eq. (S1). We can thus write

$$\ln[f_\pm(y+x,t)] = -\frac{1}{4}\int_{-\infty}^{+\infty}\frac{d\omega}{\omega}\int_{-\infty}^{+\infty}\frac{d\omega'}{\omega'}\delta(\omega+\omega')\Big\{A_1(y,t)S_x^{++}(\omega) + A_2(y,t)S_x^{-+}(\omega) \\ + A_3(y,t)S_x^{+-}(\omega) + A_4(y,t)S_x^{--}(\omega)\Big\}, \tag{S30}$$

with

$$\begin{aligned}
A_1(y,t) &= e^{-i\omega(t-t_c^y)}e^{i\omega' t_c^y} \pm e^{-i\omega(t-t_c^y)}e^{i\omega' t_n^y} \pm e^{-i\omega(t-t_n^y)}e^{i\omega' t_c^y} + e^{-i\omega(t-t_n^y)}e^{i\omega' t_n^y}\\
&\quad -\frac{1}{2}e^{-i\omega(t-t_c^y)}e^{-i\omega'(t-t_c^y)} - \frac{1}{2}e^{-i\omega(t-t_n^y)}e^{-i\omega'(t-t_n^y)} - \frac{1}{2}e^{i\omega t_c^y}e^{i\omega' t_c^y} - \frac{1}{2}e^{i\omega t_n^y}e^{i\omega' t_n^y}\\
&\quad \mp\frac{1}{2}e^{-i\omega(t-t_c^y)}e^{-i\omega'(t-t_n^y)} \mp \frac{1}{2}e^{-i\omega(t-t_n^y)}e^{-i\omega'(t-t_c^y)} \mp \frac{1}{2}e^{i\omega t_c^y}e^{i\omega' t_n^y} \mp \frac{1}{2}e^{i\omega t_n^y}e^{i\omega' t_c^y}\\
A_2(y,t) &= e^{-i\omega(t-t_c^y)}e^{i\omega' t_c^y} \pm e^{-i\omega(t-t_c^y)}e^{i\omega' t_n^y} \mp e^{-i\omega(t-t_n^y)}e^{i\omega' t_c^y} - e^{-i\omega(t-t_n^y)}e^{i\omega' t_n^y}\\
&\quad -\frac{1}{2}e^{-i\omega(t-t_c^y)}e^{-i\omega'(t-t_c^y)} \pm \frac{1}{2}e^{-i\omega(t-t_n^y)}e^{-i\omega'(t-t_c^y)} \mp \frac{1}{2}e^{-i\omega(t-t_n^y)}e^{-i\omega'(t-t_n^y)}\\
&\quad +\frac{1}{2}e^{-i\omega(t-t_n^y)}e^{-i\omega'(t-t_n^y)} - \frac{1}{2}e^{i\omega t_c^y}e^{i\omega' t_c^y} + \frac{1}{2}e^{i\omega t_n^y}e^{i\omega' t_c^y} \mp \frac{1}{2}e^{i\omega t_c^y}e^{i\omega' t_n^y} + \frac{1}{2}e^{i\omega t_n^y}e^{i\omega' t_n^y}\\
A_3(y,t) &= e^{-i\omega(t-t_c^y)}e^{i\omega' t_c^y} \mp e^{-i\omega(t-t_c^y)}e^{i\omega' t_n^y} \pm e^{-i\omega(t-t_n^y)}e^{i\omega' t_c^y} - e^{-i\omega(t-t_n^y)}e^{i\omega' t_n^y}\\
&\quad -\frac{1}{2}e^{-i\omega(t-t_c^y)}e^{-i\omega'(t-t_c^y)} \pm \frac{1}{2}e^{-i\omega(t-t_c^y)}e^{-i\omega'(t-t_n^y)} \mp \frac{1}{2}e^{-i\omega(t-t_n^y)}e^{-i\omega'(t-t_c^y)}\\
&\quad +\frac{1}{2}e^{-i\omega(t-t_n^y)}e^{-i\omega'(t-t_n^y)} - \frac{1}{2}e^{i\omega t_c^y}e^{i\omega' t_c^y} + \frac{1}{2}e^{i\omega t_c^y}e^{i\omega' t_n^y} \mp \frac{1}{2}e^{i\omega t_n^y}e^{i\omega' t_c^y} + \frac{1}{2}e^{i\omega t_n^y}e^{i\omega' t_n^y}\\
A_4(y,t) &= e^{-i\omega(t-t_c^y)}e^{i\omega' t_c^y} \mp e^{-i\omega(t-t_c^y)}e^{i\omega' t_n^y} \mp e^{-i\omega(t-t_n^y)}e^{i\omega' t_c^y} + e^{-i\omega(t-t_n^y)}e^{i\omega' t_n^y}\\
&\quad -\frac{1}{2}e^{-i\omega(t-t_c^y)}e^{-i\omega'(t-t_c^y)} - \frac{1}{2}e^{-i\omega(t-t_n^y)}e^{-i\omega'(t-t_n^y)} - \frac{1}{2}e^{i\omega t_c^y}e^{i\omega' t_c^y} - \frac{1}{2}e^{i\omega t_n^y}e^{i\omega' t_n^y}\\
&\quad \pm\frac{1}{2}e^{-i\omega(t-t_c^y)}e^{-i\omega'(t-t_n^y)} \pm \frac{1}{2}e^{-i\omega(t-t_n^y)}e^{-i\omega'(t-t_c^y)} \pm \frac{1}{2}e^{i\omega t_c^y}e^{i\omega' t_n^y} \pm \frac{1}{2}e^{i\omega t_n^y}e^{i\omega' t_c^y}.
\end{aligned} \tag{S31}$$

Using the properties of the $\delta(\omega)$ function we get

$$\ln[f_\pm(y+x,t)] = \frac{1}{4}\int_{-\infty}^{+\infty}\frac{d\omega}{\omega^2}\Big\{A_1(y,t)S_x^{++}(\omega) + A_2(y,t)S_x^{-+}(\omega) + A_3(y,t)S_x^{+-}(\omega) + A_4(y,t)S_x^{--}(\omega)\Big\}, \tag{S32}$$

where the $A_i(y,t)$ coefficients have considerably simplified to

$$\begin{aligned}
A_1(y,t) &= 2[1 \pm \cos(\omega(t_n^y - t_c^y))][e^{-i\omega t} - 1]\\
A_2(y,t) &= \mp 2i\sin(\omega(t_n^y - t_c^y))[e^{-i\omega t} - 1]\\
A_3(y,t) &= \pm 2i\sin(\omega(t_n^y - t_c^y))[e^{-i\omega t} - 1]\\
A_4(y,t) &= 2[1 \mp \cos(\omega(t_n^y - t_c^y))][e^{-i\omega t} - 1].
\end{aligned} \tag{S33}$$

Finally we arrive at

$$
\ln[f_\pm(y+x,t)] = \frac{1}{2}\int_{-\infty}^{+\infty}\frac{d\omega}{\omega^2}(e^{-i\omega t}-1)\Big\{ S_x^{++}(\omega)+S_x^{--}(\omega)\pm\cos[\omega(t_n^y-t_c^y)][S_x^{++}(\omega)-S_x^{--}(\omega)]
$$
$$
\pm i\sin[\omega(t_n^y-t_c^y)][S_x^{+-}(\omega)-S_x^{-+}(\omega)]\Big\}.
$$
(S34)

Using Eqs. (S5) and (S7), we can identify the term into brackets with $2S_{y+x}^{\pm\pm}(\omega)$ discussed in the main text, see Eq. (4). This result emphasizes that one cannot write the electronic distributions in $x+y$ only from the autocorrelations in $x$: the knowledge of the crosscorrelations in $x$ is essential. The distribution function in real time is thus given by

$$
\langle\psi_\pm^\dagger(y+x,t)\psi_\pm(y+x,0)\rangle = \frac{1}{2\pi\delta}e^{g_\pm(y+x,t)}
$$
(S35)

with the exponent

$$
g_\pm(y+x,t) = \int_{-\infty}^{+\infty}d\omega\frac{e^{-i\omega t}-1}{\omega^2}S_{y+x}^{\pm\pm}(\omega).
$$
(S36)

Recall that if we require the energy dependence of the distribution function $f_\pm(y+x,E)$ for the channel $\pm$ we need to Fourier transform the obtained solution:

$$
f_\pm(y+x,E) = \int_{-\infty}^{+\infty}dt\, e^{iEt}\langle\psi_\pm^\dagger(y+x,t)\psi_\pm(y+x,0)\rangle.
$$
(S37)

### Electron distributions

Here we sketch the simple solution of the case with $y=0$, which depends only on the evolution in the region of lenght $x$ assuming that the source terminals are kept at two different temperatures. We assume that the emission spectral noise of the $\alpha$-contact is $S_0^{\alpha\alpha}(\omega)$. Finally we also assume normal metal Ohmic contact, at the two channel boundaries, the spectral functions at inverse temperature $\beta_\alpha$ are given by Eq. (S3). When the two channel are in equilibrium $\beta_\pm=\beta$ we can easily recover the standard two-point bosonic Green function.

We find for the exponential function

$$
W^{(\alpha)}(x,t) = \frac{1}{2}\left(\sum_{\alpha=\pm}W_{\beta_\alpha}(t)+\frac{\alpha}{2}(W_{\beta_\alpha}(t+\delta t^x)-W_{\beta_\alpha}(\delta t^x)+W_{\beta_\alpha}(t-\delta t^x)-W_{\beta_\alpha}(-\delta t^x))\right),
$$
(S38)

where $\delta t^x = t_n^x-t_c^x$, which exponentiated becomes

$$
\langle\psi_\pm^\dagger(x,t)\psi_\pm(x,0)\rangle = \frac{1}{2\pi\delta}\sqrt{G_{\beta_+}(t)G_{\beta_-}(t)}\sqrt[4]{\frac{G_{\beta_\pm}(t+\delta t^x)\,G_{\beta_\pm}(t-\delta t^x)}{G_{\beta_\mp}(t+\delta t^x)\,G_{\beta_\mp}(t-\delta t^x)}}\sqrt[4]{\frac{G_{\beta_\mp}(+\delta t^x)\,G_{\beta_\mp}(-\delta t^x)}{G_{\beta_\pm}(+\delta t^x)\,G_{\beta_\pm}(-\delta t^x)}}.
$$
(S39)

Here we expressed the real time electron two point Green's function in terms of the free electron Green function at finite temperature $G_\beta(t)$. It is useful to note that, since $G_\beta(t\to 0)\to 1$, we easily recover $\langle\psi_\pm^\dagger(x,t)\psi_\pm(x,0)\rangle\to G_{\beta_\pm}(t)$ in the non-interacting limit $K\to 1$ ($\delta t^x\to 0$), as required by consistency of the theory with the previous results.

### Long distance limit

Let us consider the asymptotic behavior of Eq. (S39) in the limit $x\to\infty$. We recall that $\delta t^x=t_n^x-t_c^x=x(1-K)/v_F$, therefore in the limit $x\to\infty$ the two ratios within fourth roots in Eq. (S39) cancel and we obtain

$$
\lim_{x\to\infty}\langle\psi_\pm^\dagger(x,t)\psi_\pm(x,0)\rangle\to\frac{1}{2\pi\delta}\sqrt{G_{\beta_+}(t)G_{\beta_-}(t)}.
$$
(S40)

Using the expression:

$$
G_\beta(t) = \frac{\left|\Gamma(1+\frac{1}{\beta\omega_c}+i\frac{t}{\beta})\right|^2}{\Gamma(1+\frac{1}{\beta\omega_c})^2(1+i\omega_c t)}
$$
(S41)

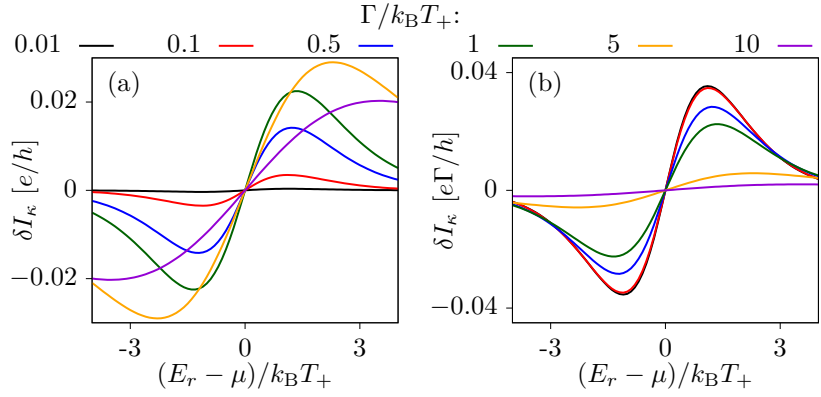

FIG. S3. (a) Difference $\delta I_\kappa = I(\kappa=0) - I(\kappa=1)$ of the currents for channels that after interacting along a distance $\tilde{x}$, have their crosscorrelations reset ($\kappa = 1$) or not ($\kappa = 0$) before interacting along a distance $\tilde{y} = 2x$, with $\mathcal{T} = 0.5$. The scatterer is a finite width antidot with a resonant transmission $\mathcal{T}(E) = (\Gamma/2)^2/[(E - E_r)^2 + (\Gamma/2)^2]$, where $E_r$ is the resonant energy. (b) The same curves, normalized by the width $\Gamma$, for comparison with Fig. 4(d) in the main text (note that there we had $\mathcal{T} = 10^{-4}$). Very narrow resonances are more precise (b), but give much smaller signals (a).

and introducing the notation

$$G_\beta^{(\nu)}(t) = (G_\beta(t))^\nu \tag{S42}$$

we can write the energy distribution function as

$$f(E) = \frac{1}{2\pi\delta} \int_{-\infty}^{+\infty} dt\, e^{iEt} G_{\beta_+}^{(1/2)}(t) G_{\beta_-}^{(1/2)}(t). \tag{S43}$$

Using the Fourier transform

$$G_\beta^{(\nu)}(t) = \int_{-\infty}^{+\infty} \frac{d\epsilon}{2\pi} e^{-i\epsilon t} G_\beta^{(\nu)}(\epsilon) \tag{S44}$$

we finally write

$$f(E) = \frac{1}{2\pi\delta} \int_{-\infty}^{+\infty} \frac{d\epsilon}{2\pi} G_{\beta_+}^{(1/2)}(\epsilon) G_{\beta_-}^{(1/2)}(E - \epsilon). \tag{S45}$$

We now recall that in the scaling limit $\beta_\alpha \omega_c \gg 1$

$$G_{\beta_\alpha}^{(\nu)}(\epsilon) = \mathcal{D}_\alpha^{(\nu)}(\epsilon) f_\alpha(\epsilon), \tag{S46}$$

with the equilibrium Fermi distribution $f_\alpha(\epsilon)$ and the interacting density of states

$$\mathcal{D}_\alpha^{(\nu)}(\epsilon) = \frac{(2\pi)^\nu}{\omega_c \Gamma(\nu)} \left(\frac{1}{\beta_\alpha \omega_c}\right)^{\nu-1} \frac{\left|\Gamma\left(\frac{\nu}{2} + i\frac{\beta_\alpha \epsilon}{2\pi}\right)\right|^2}{\left|\Gamma\left(\frac{1}{2} + i\frac{\beta_\alpha \epsilon}{2\pi}\right)\right|^2}. \tag{S47}$$

## DETECTION WITH A FINITE WIDTH RESONANCE

In the main text, we consider an antidot with a very narrow transmission, $\mathcal{T}(E) = \Gamma\delta(E - E_r)$, which gives a nice energy resolution of the distribution but results in tiny currents. In a real experiment, the width of the resonance needs to be taken into account. This is what we do here, assuming a Breit-Wigner scattering amplitude [S3] giving the transmission probability:

$$\mathcal{T}(E) = \frac{(\Gamma/2)^2}{(E - E_r)^2 + (\Gamma/2)^2}. \tag{S48}$$

As shown in Fig. S3, considerably large widths of the order of the thermal energy $k_{\mathrm{B}}T_+$ are still sensitive to the effect of the crosscorrelations, $\delta I_\kappa$, while at the same time being easier to detect.

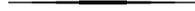



\end{document}